# Public Technologies Transforming Work of the Public and the Public Sector

Seyun Kim
seyunkim@cs.cmu.edu
Carnegie Mellon University
Pittsburgh, Pennsylvania, USA

Bonnie Fan
byfan@andrew.cmu.edu
Carnegie Mellon University
Pittsburgh, Pennsylvania, USA

Willa Yunqi Yang
yunqiy@uchicago.edu
University of Chicago
Chicago, Illinois, USA

Jessie Ramey
jramey@chatham.edu
Chatham University
Pittsburgh, Pennsylvania, USA

Sarah E Fox
sarahfox@cmu.edu
Carnegie Mellon University
Pittsburgh, Pennsylvania, USA

Haiyi Zhu
haiyiz@cs.cmu.edu
Carnegie Mellon University
Pittsburgh, Pennsylvania, USA

John Zimmerman
johnz@cs.cmu.edu
Carnegie Mellon University
Pittsburgh, Pennsylvania, USA

Motahhare Eslami
meslami@andrew.cmu.edu
Carnegie Mellon University
Pittsburgh, Pennsylvania, USA

## ABSTRACT

Technologies adopted by the public sector have transformed the work practices of employees in public agencies by creating different means of communication and decision-making. Although much of the recent research in the future of work domain has concentrated on the effects of technological advancements on public sector employees, the influence on work practices of external stakeholders engaging with this sector remains under-explored. In this paper, we focus on a digital platform called OneStop which is deployed by several building departments across the U.S. and aims to integrate various steps and services into a single point of online contact between public sector employees and the public. Drawing on semi-structured interviews with 22 stakeholders, including local business owners, experts involved in the construction process, community representatives, and building department employees, we investigate how this technology transition has impacted the work of these different stakeholders. We observe a multifaceted perspective and experience caused by the adoption of OneStop. OneStop exacerbated inequitable practices for local business owners due to a lack of face-to-face interactions with the department employees. For the public sector employees, OneStop standardized the work practices, representing the building department's priorities and values. Based on our findings, we discuss tensions around standardization, equality, and equity in technology transition, as well as design implications for equitable practices in the public sector.

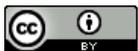

## 1 INTRODUCTION

Public sector agencies deploy and adopt technologies to improve efficiency and make use of limited social service resources. These technologies, including digital portals, data-driven and algorithmic innovations, are transforming the landscape of work within the public sector by significantly shifting employees' work practices [2, 34, 35, 40, 42]. Recent research has shown how public technologies have shifted and impacted the employees' daily decision-making practices in high-stake scenarios such as child-welfare [42], recidivism [6] and homelessness [40, 44]. Public sector agencies have adopted online platforms as means of communicating and sharing data among employees, creating digitization of in-person interactions [13, 21, 66].

However, public sector employees are not the only group influenced by public technologies in their work practices. In response to the COVID-19 pandemic, several cities across the U.S. have made significant efforts to enhance their data infrastructure and service delivery through online portals such as OneStop [24, 58, 60, 68]. This portal aims to transition in-person communications to an online format and consolidate various steps and services into a single point of contact between the public sector and residents. One of the sectors to adopt this technology has been municipal building departments, whose responsibilities include issuing permits and licenses, as well as conducting necessary inspections for businesses.

While this technological shift can affect the employees' work practices in building departments, its impact extends to the public (e.g., business owners, architects, developers, and other workers) who interacts with the public sector. However, the extent to which this transition impacts the work of these constituents remains unclear.

We aim to understand how the technological shifts in the public sector impact work practices for both public sector employees and the residents interacting with the services. In particular, we investigate several work contexts that are impacted by the adoption of OneStop development in a municipal building department at a Northeastern mid-sized U.S. city (hereafter, the "City"): 1) small, local, women, and minority-owned businesses that require permits, licenses, and inspections to launch their new ventures, 2) experts involved in the construction process (e.g. architects, contractors, developers), and 3) community representatives who understand the building department's impact to the neighborhood and business owners, and 4) building department employees who issue permits and licenses, schedule, and execute inspections.

We found that local business owners, particularly "first-time applicants", experienced inequitable knowledge and financial access in developing their businesses. The transition to OneStop exacerbated inequities as face-to-face interactions between employees and constituents reduced and created frustration. Participants spoke of their fears around human bias and discrimination, calling out issues of race and gender. These fears impacted the way they chose to engage with the building department. Finally, the building department shared that staffing shortages, as well as a culture change driven by the new kind of staff being hired and the roll-out of the online service, all impacted service delivery. The changes seemed especially bad for local businesses that suffered more from errors, uncertainty, and delays.

Our findings also surfaced the building departments' managerial priorities and values, emphasizing the standardization of service delivery. This priority for standardization was reflected in the transition to OneStop. The technology transition not only shaped the characteristics of the labor force decreasing opportunities for merit-based vertical promotions but also created a culture of supervisors monitoring worker productivity. When comparing the constituents' expectations of the building department's role and the managerial priorities, we found a critical misalignment. While constituents expected the building department to address all of their needs, the department prioritized meeting department metrics for delivering public services.

Our work contributes to HCI detailing how public sector technologies impact the work for not only public sector employees but also diverse constituents in the context of municipal building departments. We detail the tensions surrounding the technology that standardizes work and the challenges inherent in meeting the broad needs of municipal public service constituents.

## 2 RELATED WORK

### 2.1 Public Sector Technology Impacting the Workplace

Technologies, including algorithms, data-driven innovations, and digital platforms are rewriting the rules of work in the public sector for government employees. Prior work has investigated technologies transforming the public sector employee's professional identities, decision-making process, and bureaucratic structure [45, 53].

Past literature has explored how technological innovations transform the public sector's vertical hierarchy and bureaucracy [66]. Communication channels in the public sector, known to be rigid and inflexible, transformed to a more fluid and spontaneous process, creating a horizontal relationship among employees [63, 66]. In addition, data sharing across sub-departments and among employees created easier access to information and a non-hierarchical work culture [66]. Technology has also transformed how the public sector communicates with its constituents. Constituents complete electronic forms when submitting paperwork to the public sector, reducing face-to-face interactions [9]. As residents access information through the government's online portals, employees' professional identities have been reshaped. With the growth of the internet and digital support, what was once the main form of support transitioned to one of many support that residents could receive [66].

Public sector employees' daily decision-making processes have been affected by automated systems integrated into the workplace [73, 74, 81]. A body of work investigated algorithms used for managerial decisions, either replacing managers or informing decisions [57, 77]. In addition, public sector employees expressed concerns about automated systems reducing their discretionary power [4, 18] and creating inconsistent decisions that may lead to harm [11, 75]. Researchers explored how the design of these systems created tensions between the public sector employees' agency and the transactional standardization that computer systems impose [65]. While street-level bureaucrats, who are on-the-ground employees, make reflexive decisions, algorithms lack the flexibility to make in-the-moment decisions [2, 3, 48, 62]. Møller et al. investigated how employees' knowledge, expertise, and professional judgment interplay with the technology's capacity to process large amounts of data and identify patterns that may exceed an individual's experience [34]. Kawakami et al. elaborated that frontline workers in a child welfare agency rely on algorithmic decision support tools while considering the tools' capabilities and limitations, organizational pressure, as well as discrepancies between their own judgment and the algorithm's decision [42]. Ammitzbøll Flügge et al. discussed how caseworkers considered automated systems helpful as a support tool when they need extra assistance to advocate their case to management or receive advice [3]. Despite the attention to technology's impact on public sector employees, there's limited research on its effects on the work of public service constituents.

### 2.2 Data and Technology in Urbanism

Municipal building departments play a critical role in setting the stage for a safely built environment by overseeing the construction and development of urban projects. Data analytics and technologies have become critical tools for providing new insights for cities to see the overall trend of urban development. With the importance of data monitoring city functions, researchers have investigated the topic of urban informatics. Urban informatics is defined as the "study, design, and practice of urban experiences [...] that are created by new opportunities of real-time, ubiquitous technology" [25]. Researchers have explored how to develop a sustainable neighborhood with the emergence of new technologies and data [1]. Ortegón et al. investigated the use of digitization systems by the Colombian government to analyze land use plans and urban growth [61]. Zuniga et al. explored how urban logistics, which focused on route optimizations and land use planning, helped create a city that meets residents' needs and ensures quality of life [50].

When adopting and integrating technologies in cities, researchers have investigated the importance of residents' quality of life and the neighborhoods. Williams et al. raised concerns about urban informatics that considers cities only as an "economically and spatially social form". Instead, cities should be examined from the perspective of the residents' experiences living in the city [84]. Freeman et al. argued that when implementing technologies in cities, it is crucial to consider the residents' perceptions and attitudes toward these technologies. Technologies should be integrated within the urban environment without disrupting the neighborhoods' unique characteristics [28]. To do so, Freeman et al. emphasized the need for a broader representation of urban inhabitants, especially those from Western cities, in the design of urban technologies [27]. A concrete example where a city's infrastructure did not meet the community's needs is illustrated by Joshi et al., where the city had focused on creating a "smart" water infrastructure before considering its functionality. Therefore, the authors emphasized that the geographical and cultural context of the city is critical information needed when adopting new technological advancements [39]. Prior work discussed how technologies and data integrated in the public sector impact the residents' lives. Building upon prior work investigating public technologies in the context of urban development and building departments, our work explores how constituents of a municipal building department are affected by technologies, particularly in work practices.

## 3 BACKGROUND
### 3.1 The Context of the Building Department

The City's building department is responsible for keeping the built environment safe. This involves issuing permits and building licenses, inspecting buildings and new constructions, and enforcing building codes. The building department is one of many players in urban development that impact the quality of life in city neighborhoods as it is responsible for the demolitions of buildings or the repair of built structures that violate safety regulations [7, 22, 43, 59]. The constituents with which the building department interacts include residents who aim to develop a commercial enterprise in the local community as well as non-residents, such as architects, developers with expertise in construction and maintenance of built environment.

### 3.2 The Tale of Two Businesses

Our work is motivated by the tale of two businesses representing the small business owners' experiences with the City's building department and how the interactions impacted the owners' work practices and businesses. A married couple planned to open a coffee shop in a neglected neighborhood. However, while interacting with the building department, the couple realized that obtaining the authorization for their occupancy permit would take many months, which could potentially delay their opening day. Additionally, the couple only later realized that their permit applications were denied. The lack of transparency in the permitting process motivated the couple to contact their city councilperson to intervene, which resulted in the approval of their permit application without additional costs or delays [69]. A similar story of a young entrepreneur starting a bakery did not have the same happy outcome. This story came to light for us during a personal conversation with a city councilperson. While submitting the occupancy permit, the entrepreneur only later realized that her shop required an expensive modification to satisfy the safety code required by the City's building department. She could not afford the construction cost and had to abandon her business plans. These two anecdotes highlight how obtaining a permit and interacting with the building department significantly impacted whether these entrepreneurs could start their businesses. Despite the high-stakes impact on small business owners, it is unclear how the building department's transition to technology influenced the work of these local business owners.

### 3.3 Technology Transition: OneStop and Data Collection

As part of a major push to advance the City's infrastructure, the building department underwent a technology transition that influenced the department's service delivery model. During the COVID-19 pandemic, the City launched a web portal called OneStop, which was meant to be a single point of digital contact between residents and City departments. The building department was the first to participate in this initiative. OneStop is a software product of an enterprise system called Accela [1] that specializes in providing government software solutions for permitting and licensing. The building department had been using the internal mode of this system for a decade before the launch of OneStop. Once OneStop was launched, applicants no longer went in person to the building department to file their building permits. Digitization of the application process enabled the building department to incorporate data analytics into its planning and process improvement initiatives.

### 3.4 A Typical Workflow of the Building Department

A building permit is required for any construction when starting a business. This process involves communicating with the department and completing the permit application through OneStop. After the building department receives the completed permit application, certified inspectors evaluate the design and construction of the business owner's facilities based on building codes, a set of standards to ensure safe building structures. The permit is accepted when all inspections are approved. This workflow often involves multiple stakeholders. Business owners may need to hire experts such as architects and contractors or contact registered community organizations for their expertise and communication with the building department.

---

[1] https://www.accela.com/

## 4 METHOD

We conducted 22 semi-structured interviews with employees from the City's building department and their constituents. In this section, we describe our interview process, participants, and our approach to analyzing the collected data.

### 4.1 Study Design

To gain insight into the City's context and shape the interview protocols, we attended the City's urban development community meetings and navigated the details of the permit application process. As a result, we developed two interview protocols: one for constituents who interact with the building department and one for the building department employees. Constituents were drawn from a) local business owners, b) experts involved in the construction process (e.g., architects, contractors, developers), and c) community representatives. Municipal employees were drawn from multiple sub-departments, roles, and hierarchies. Here, we illustrate the different topics asked of each stakeholder:

- Local business owners & Experts: 1) We investigated the constituents' experiences while interacting with the building department. These experiences include roadblocks and the duration of the application process; 2) We probed existing resources or support from the building department, including flexibility in the application process; 3) We asked about the business owners' ties to the neighborhood to gain insight into the owners' connections to the community; 5) We asked about changes or desires that the participant would like to see in the building department.
- Community representatives: 1) We focused on understanding the broader patterns of interactions between local business owners and the building department; 2) We asked how local businesses influence the quality of life in neighborhoods; 3) We probed how the building department's processes impact the City's neighborhoods.
- Building department employees: 1) We asked about the protocol of building department work practices, decision-making processes, and insights regarding where constituents need support; 2) We probed about experiences interacting with constituents, as well as roadblocks constituents experience; 3) We asked about priorities and broader goals regarding how applications are processed and revised; 4) We investigated the employees' standards of success and how work is being evaluated.

After the study, we contacted participants identified as constituents to complete an optional post-survey questionnaire asking for demographic information and the frequency of interactions with the building department.

### 4.2 Recruitment and Gaining Access: Challenges and Lessons Learned

We recruited local business owners through business support organizations that provide training and resources for entrepreneurs to start their businesses (e.g., women's entrepreneurial center). In addition, we communicated with business owners via phone calls and emails. We recruited experts (e.g., architects, developers) and community representatives through emails, phone calls, or word of mouth.

As we recruited and interviewed constituents, we worked to establish a collaborative relationship with the City's building department. Although the department leadership was interested in partnering with the research team to investigate improvements in the permit application process, much time and effort was needed to establish trust and clarify legal requirements. We shared the initial findings from the first few constituent interviews with the department's leadership to gain buy-in. Our relationship with the leadership facilitated the connection with the building department managers, who sent our recruitment messages and flyers to their employees. These employees have direct interactions with the constituents who apply for the permit.

The interview protocol covered sensitive topics related to department processes, including inequitable practices. We were concerned that public sector employees might hesitate to participate in the interview due to fears of potentially damaging the relationship with their supervisors and the department. We discussed this recruitment challenge with the Carnegie Mellon University Institutional Review Board, which suggested that we refrain from asking public sector employees participating in our study for their demographic information to reduce the risk of being identifiable. Furthermore, the recruitment process was conducted through a Google form to ensure that the employees' agreement to participate in our study was not traceable through their organization email.

### 4.3 Participants

We interviewed 22 participants. The participants consisted of 13 constituents, including 5 local business owners (*B#*), 2 local architects (*A#*) and 2 local contractors (*C#*) who frequently function as intermediaries between the building department and a business owner; 1 developer (*D#*) who runs projects involving many businesses, and 3 community representatives (*CM#*) who advocate for change around urban development to improve the quality of life in the City. Table 1 summarizes the demographic information of the 13 constituents. We interviewed 9 department employees(*DE#*), including 4 with management responsibilities and 5 with staff roles. The study was carried out over the phone or via video conferencing (e.g., Zoom) for 60 minutes, with a compensation of a $25 gift card per hour. With the consent of the participants, the audio was recorded and transcribed.

### 4.4 Analysis

To understand the participants' responses to the interviews, we created customer journey maps [85], which detail how the constituents' experiences change as they move through various touch points and goals. These insights reveal where constituents are satisfied and identify pain points that help improve services. In addition to customer journey maps, we created service blueprints to capture the interaction between the front-stage and back-stage processes of the building department's permit application processes [67].

The research team conducted interpretation sessions [8] using an inquiry process to draw insights from each interview. The interpretation session, with three authors, involved addressing ambiguities or disagreements in the interview analysis. We performed an affinity diagramming process [8] from the insights derived from the interpretation sessions. We refined and created the affinity diagram focusing on the constituents' experiences and work practices, the building department's business processes, and the impact of the technology transition.

| Participant | Gender | Ethnicity | Interaction with the building department |
| --- | --- | --- | --- |
| B1 | F | - | - |
| B2 | M | White | 1 commercial building permit |
| B3 | F | African American | Started but did not submit a permit application |
| B4 | F | - | - |
| B5 | M | Black | Did not submit a permit application |
| C1 | M | White | 6-10 residential building permits per year |
| C2 | M | White | More than 3 commercial building permit applications |
| D1 | M | Black | More than 3 commercial building permit applications |
| CM1 | F | White | NA |
| CM2 | M | White | NA |
| CM3 | F | Black | NA |
| A1 | M | White | - |
| A2 | M | - | - |

**Table 1: The table includes only *the constituents*. Building department employees were not asked to provide demographic information to protect their identities. The optional post-study survey consisted of demographic information and the frequency of interaction with the building department. "-" means that the participant did not respond. "NA" indicates that the question was not applicable to community representatives.**

## 5 FINDINGS

Our findings reveal barriers and misalignment in interactions between the building department and its constituents. The introduction of OneStop exacerbated challenges and unintentionally created new equity barriers. This section details common barriers, such as constituents' lack of experiences with the building department and limited financial resources. Small, local business owners also had to navigate perceived discrimination while interacting with the building department. We show how implementing OneStop as a digital service platform not only reduced two-way communication between the building department and its constituents but also exacerbated existing inequities. We then investigate the department's priorities for efficiency and standardization, and show how these priorities influence data and technology use as well as management of bureaucratic labor. Finally, we discuss the misalignment between constituents' expectations and the department's role. We end by connecting insights from our interviews to the wider city government and politics.

### 5.1 Knowledge and Financial Barriers, and the Fear of Inequitable Practices

High barrier to entry made the process especially challenging to local residents aspiring to open a business. Building relationships with experts and the building department became a necessity to overcome knowledge barriers. For small business owners, financial needs made delays in the permitting process personally costly. Small business owners, especially those identified as minority groups, expressed fear of bias and addressed the corresponding coping strategies.

We use a few terms to distinguish the types of constituents that the building department serves. "First-time applicants" have little knowledge or experience in the permitting process, while "frequent flyers" are identified as those with expertise. All local business owners started as "first-time applicants", some gaining more experience to "frequent flyers" status, while developers and architects were defaulted "frequent flyers" of the construction process.

*5.1.1 High Expertise Barriers for First Timers Create an Inequitable Knowledge Economy.* "First-time applicants" experienced challenges in navigating the permitting process and using OneStop compared to "frequent flyers" who already had experience interacting with the building department. Three participants who were "first-time applicants" during their interactions with the building department expressed confusion and frustration about the information gathering process. For example, B1, a "first-time applicant", shared the challenges of her initial information gathering process that occurred in person and over the phone prior to the roll-out of OneStop. She explained the difficulties in getting access to the building department representatives: *"I was trying to find a phone number to call, so that I could talk to a human being and express my [...]concerns, and it was hard to find a phone number. [...] It was like a goose chase to try to get a person on the phone"* (B1). She communicated with a public sector employee a few times, clarifying the materials she needed over the phone. However, when the participant visited the building department to file the application, she had a different experience from her phone calls: *"I will not forget that person that was at the desk [...] When we finally got through the line, [the employee] yelled at us because we did not have everything that [the employee] needed. [...] They were like yelling at us [...] trying to make us feel like stupid and small"* (B1). For "first-time applicants", this knowledge barrier necessitates relationship building with an expert. However, for "frequent flyers", they often identified and sought out contacts in the department that were easier to work with (B4, C2).

Another knowledge barrier involves the need to interpret the building code and gain the expertise in the permitting process. An architect (A2) finds that very few people know the building code in detail, with only one code expert in the building department with whom constituents can communicate. Different interpretations of building codes resulted in delays, either in the design approval stage or in receiving contradictory determinations in inspections.

"First-time applicants" lack experience, preparation, and personal workarounds making them susceptible to disadvantaged scenarios. Disadvantaged scenarios involve being told that changes must be made in a previously approved application material (B2), as well as meeting building requirements that are arduous for business owners, potentially leading to delays and budget constraints (B1). Inconsistencies between inspection decisions and the lack of leeway to interpret gray areas in the building code were common frustrations (C2, DE5). Navigating construction requirements is complex for all constituents, especially since each commercial project requires unique configurations and interpretations. However, this complexity disproportionately affects "first-time applicants" due to the challenges of accessing the necessary resources.

*5.1.2 Financial Barriers and Disproportionate Impacts.* Access to financial resources and the ability to withstand delays created vastly different stakes for local business owners, particularly "first-time applicants". Local business owners, particularly "first-time applicants", struggled to identify and secure financing options. To secure finances, business owners would run crowdfunding campaigns (B1) or obtain loans (A2, B3). An equity-focused community representative revealed the elitism involving the funding process that supports businesses (CM3). Registered community organizations are key in connecting neighborhood residents with potential financial opportunities (CM1); however, market constraints often influence which projects receive support (CM2).

Local business owners expressed frustration about making unanticipated personal investments in their business due to the building department's inconsistencies or arbitrary actions. Unexpected delays or expensive changes to plans impacted small, local, "first-time applicants" the most due to their financial vulnerability: *"I went ballistic because [...] we've already spent almost all of our budget. [...] I said [to the building department that] you will have to pay for it because I'm going to sue you. Or you need to recognize that you made a mistake on something that really isn't that big of a deal"* (B2).

B1, a "first-time applicant", discussed the intertwined nature between financial resources and information access. She found that the building requirements to prepare her space exceeded her budget, creating a year delay and forcing her business to find a new location. In contrast to large businesses that have flexibility and abundant resources, small, local businesses suffer from unexpected delays that could lead to uncertainty in even opening the business: *"We really just ended up having to change our entire plan for the business, just to be able to be open at all. [...] [Unexpected delays and changes] kill small businesses because small businesses, you're not like a McDonald's or a Walmart where [large businesses] have expendable money where [large businesses] [...] can throw thousands of dollars wherever. Small businesses, or people that have scraped things together have probably done things like crowdfunding"* (B1).

Several participants discussed the need for accountability and leeway from the public sector when unexpected outcomes occur: *"If [the building department] had said [the required changes] when I handed [them] my engineers plans [...], I would have said, no problem. [...] But they didn't, and there's no consequence for them and 100% consequences for us."* (B2). C2 expressed frustration as the *"[Building department] had no discretion to give us any leeway with it, even though they admitted that it was their mistake"* (C2).

Lack of leeway and accountability creates a ripple effect that negatively impacts the start and development of a business. A local business owner (B4), a "frequent flyer", described the trade-off between going into business by herself and leasing from a large developer. She described how unexpected delays from the building department caused contractors to halt her project. This challenge made going through the permitting process as a small business much riskier: *"You need those contractors in line to file those papers[...] then to come back and be 'Okay, well, we really can't start anything for four months', you are going to lose that [contractor]. Unless you're a bigger developer and then you've got like you know massive contracts, but I'm just paying somebody a little over like $200,000"* (B4). This frustration reflected the disadvantage of small business owners compared to large developers, who receive support from the building department and have the flexibility to run multiple projects simultaneously.

*5.1.3 Fear of Discrimination and Coping Strategies.* Several minority business owners perceived discrimination during the permitting and construction process. B1 shared that when *"talking to other business owners [there] seems to be a consensus that [...] it is definitely made much easier for, especially white men [...] [to complete the permitting process] [...] For women owned businesses, there seems to be for sure a bias, and a lot of issues"* (B1).

Therefore, B1 purposely reached out to male contractors to avoid potential negative gender biases in communication with the building department: *"Honestly, as unfortunate as it is, we're just going to try to see if that is something that our landlord can help us with because he is like a contractor and he's a man. He's like a white man, and I feel like they will be less terrible to him"* (B1).

B1's case was the only direct encounter with potentially biased treatment identified in our interviews; however, her experiences and coping strategies were confirmed by other participants. For example, a male participant, B2, shared that although he and his wife were jointly opening their business, he would take on communications work with the building department as they had heard of difficulties women faced with the building department in being taken seriously: *"We made a very intentional decision that we know, unfortunately, that women and people of color are often not taken as seriously by mostly white mostly male bureaucratic infrastructure. [...] Because of those unconscious biases, those pernicious unconscious biases, we wanted to try and avoid that and get as clean a shot"* (B2).

A female business owner who owned several local establishments, B4, found the need to exercise constant self-advocacy and to befriend male intermediaries who would provide expertise or intervention to aid the process of construction and permitting: *"I would never want to work with like a developer, [...] just because it's* *very obvious that they've created a system that benefits other rich white men. [...] I was very fortunate to have again created a good relationship with my building inspector through that process, so he was a big help."* (B4).

Some local business owners devised partial remedies for resource barriers through community building and mentorship. As applicants found it challenging to find useful information, they had to rely on alternative ways to gather information through guidance from other experienced constituents, mentorship, and existing relationships (B1, B2, B3, B4). Local developers sometimes took on the consultant role to help minority-owned business owners interact with the building department and connect with experts such as contractors and architects: *"We've done it in the form of a consultant [...]. We've also helped by connecting [local business owners] to other practitioners, the architects or engineers"* (D1). All of our female participants learned and benefited from a women's entrepreneurship center: *"[The women's entrepreneurship center] was a great resource for support , and [...] gender specific"* (B3). However, these coping strategies to seek relationships with experts give the greatest advantage to those close to that community of experts. Access to the necessary expertise is modulated by the ability to tap into a network of these experts.

## 5.2 "OneStop Feels Like 20 Stops": Technology Transition Reducing the Two-way Communication Between the Constituents and the Department

The building department's transition to an online service changed constituents' interactions with the department, reducing the opportunity for two-way communication between the building department and constituents. Constituents who previously went in-person to the building department offices to file building permits now uploaded everything online. As a permit technician from the building department (DE6) explained, this was a significant change in data entry. Prior to OneStop, permit applications were created through a joint conversation with constituents and building department employees, who asked clarification questions about the applicants' needs and intentions.

*5.2.1 Standardization and Reduction of Two-Way Communication.* According to the department leadership, OneStop was designed to provide a standard and consistent experience (DE5). It was intended to reduce the back-and-forth interactions between the constituents and the employees that B1 experienced (Section 5.1.1). However, this reduction of direct communications also hindered the opportunity for applicants to align their plans with the departments' processes and structures. The transition to OneStop eliminated the in-person conversation that previously helped translate applicants' needs into department requirements: *"On the online portal, we're not taking an application and interpreting it. All that has to be done on the constituent which causes quite a bit of friction. Pre-pandemic, I could take a paper application and look at it and say 'Oh, [the applicants] actually kind of translated into our own jargon.' [Before OneStop] all of the data entry used to be done by us, and now all of the data entry is done by the applicant [...] a lot of our work is figuring out: 'what did [the applicants] mean?'"* (DE6).

With the burden of data entry now on the applicants, the participants described the interface of OneStop difficult to navigate even with expert knowledge. One architect commented:*"The interface asks the same questions for every single project. A lot of the time I'd say like 50 plus percent of the time they're like half of the questions that don't apply.[...] I have to make the call on whether to leave [a question] blank or type something in or try to type a little note in or something to give the person who I'm not sure is reviewing it."* (A1). Creating a frustrating experience for constituents, A2 shared that *"OneStop feels like 20 stops [...] I say 15 out of those 20 stops you never needed to do in the 1990s up through the 2013"* (A2). In the early stages of the application process, *"you don't even know who you're working with"* (A2). Hence, even experts found it frustrating that OneStop created a one-way relationship, where *"it's up to the staff to decide whether to send you an email instead of going through the OneStop system if they get frustrated with it, just as much as we do, by the way"* (A2).

*5.2.2 (Un)interpretable Transparency.* The public sector employee argued that the roll-out of OneStop increased transparency of the permitting process to constituents by providing resources and communication lines to the public. Resources such as bulletins for common questions with the department's interpretation and standard checklists (DE5) have become available online to help applicants with the process. The summaries of the inspection reports and the status of applications are available to the public and applicants in real-time (DE5, DE2). However, constituents argued that such transparency mechanisms were challenging to interpret, particularly for "first-time applicants". Public sector employees shared that technological illiteracy prevents constituents from understanding the permitting process, creating the need for "civilian usability" that reduces technical jargon (B4, DE6). "Frequent flyers" also did not find OneStop transparent enough, resulting in the creation of self-made evaluations to predict the duration of the application using publicly available status updates (A1). Public sector employees shared that constituent support tools, such as a chat system connected with 9 support staff, emails, and phones, improved communication with constituents. However, none of the constituents in our study mentioned using the chat support feature.

## 5.3 Internal Building Department Priorities and Values

While the previous sections describe the interactions between constituents and the building department through OneStop, this section investigates the department priorities and structure affecting the public service delivery. The department's priority of standardization resulted in the roll-out of OneStop, which not only made constituents' experiences applying for a permit more uniform but also generalized labor for the department employees. The work of department employees became standardized to *"a very factory-like system"* (DE4), resulting in a decrease in respect for merit in the department's bureaucratic structure. We describe managerial priorities and changes in labor with the transition of technology.

*5.3.1 Managerial Priorities.* We highlight the priorities that the leadership shared for the department, primarily based on achieving good metrics for the department's Service Level Agreement (SLA's) (DE3, DE4, DE7, DE9). SLA's are promised turnaround time windows for different permit types that are set by city ordinance. For example, for commercial structures, the SLA's were a 30 business day turnaround for permit applications. Department leadership saw themselves doing very well in achieving their SLA's performance metrics (DE3, DE4). The department also tracked other measures, such as the number of inspections, violations, and complaints. This set of priorities was implemented at all levels of the department, where the employees' work priorities focused on addressing open permit queue items and supervisors monitoring and tracking SLA's.

*5.3.2 Increasing Hierarchy, Decreasing Merit.* The shift to generalized labor was part of a standardization effort in which management hoped to reduce variations in constituents' experiences interacting with the building department. However, this standardization also meant that the employee's view of their expertise and agency was diminished and subject to correction and review.

Prior to OneStop and standardization efforts, the building department valued the individual merits of the employees. A participant in the leadership position described that the expertise in building codes and seniority influenced the promotion to the current position (DE9). Describing themselves as knowledge experts (DE5), employees saw their knowledge in the building code as a point of agency (DE7). Each inspector had a specific discipline - e.g., building, fire, mechanical, electrical (DE5). However, with efforts to standardize the permitting process, participants expressed that today the labor force transitioned from blue-collar to white-collar, where new employees are college graduates with project management degrees (C1). Promotion through individual merit became limited (DE7) as inspectors were trained to conduct a range of "combination inspections" rather than gaining one specific expertise (DE2, DE5).

*5.3.3 Department Use of Technology for Internal Management.* Elaborated in Section 3, OneStop is also an internal software tool used by the building department employees. As OneStop enabled data collection for the building department, data analysis became a tool for employees to reason and reflect on pain points. OneStop improved work productivity (DE8) through real-time communication and information sharing, helping employees upload and confirm information (DE2, DE7), and inspectors address inconsistencies in interpreting building codes (DE5, DE2). Supervisors used analytical approaches to evaluate failure and completion rates, such as "five failed commercial inspections" or "top failed residential inspections" (DE3). Employees leveraged data analysis to understand the reasons for inspection failures that could be caused by user error or wrong configurations (DE5).

Data analytics through OneStop enabled leadership to efficiently monitor worker productivity. Prior to OneStop, supervisors and managers were using *"a formula [...] kinda like swag math"*(DE9) to consider availability and productivity of the workers: *"Depending on what type of permit it is, we will assign a value [...] Also variable is the ability of the available examiner. I have some examiners that are really fast and thorough, I have some that are very fast and an accurate, I have some that are very slow and methodical"* (DE9). Hence, supervisors appreciated OneStop's ability to track worker productivity, as the analysis could be used to examine worker management strategies such as budgeting, workload, and hiring (DE9). In the inspection process, leadership rearranged inspectors' workload to ensure that SLA's are met for the constituents (DE5). While leadership applauded the analysis feature of OneStop (DE4), support staff noticed the growing documentation work and monitoring (DE3). The implementation of data and technology highlights conflicting desires between leadership and employees. As one support staff put it: *"Our supervisors love it [but] I see us going in the opposite direction of what the staff would like [...] staff might want less thorough, but I don't ever see us going back. [...] It's tough but our leadership is on the same page"*(DE3).

## 5.4 Misalignment Between the Building Department and Constituents

We identified a misalignment between the constituents' expectations and the building department's mission. The constituents expected two-way communication and accountability for errors and delays from the building department. They expected the department to address the constituents' needs, including zoning issues which are outside of the department's responsibilities. On the other hand, the department prioritized hitting its metrics for service delivery.

*5.4.1 Misalignment of Expectations and Department's Role.* Constituents expected direct support from the building department in navigating the permitting process. However, the building department focused on compliance and efficiency. The roll-out of OneStop exacerbated this misalignment. Due to the barriers to completing the permit application process illustrated in Section 5.1.1 and Section 5.2.1, constituents expected that the department would play a more supportive role, one that wants to work with small businesses rather than being an *"unserious partner"* (B2). Constituents wanted additional explanations and support in understanding the permit requirements and rationale (DE6). As the constituent's need for support was not satisfied, complaints about the building department were brought through council members or the 311 response center (DE7, DE8). However, despite the constituents' expectation for greater support from the building department, the department leadership applauded the transition to OneStop creating standardization of services and meeting SLA's (Section 5.3.1).

Constituents considered the building department the "final gatekeeper" ensuring compliance with safety and building codes (DE4). However, this expectation of the building department's role did not align with the constituents' actual experiences. Constituents faced frustration in interacting with other urban development departments such as zoning or planning, due to the lack of inter-agency communication with the building department (C2, B4, B1). As constituents ping-ponged between these different urban development departments, creating slowdowns in the application process, the building department bears the brunt of the frustration (A1, A2).

*5.4.2 Relationship to Wider City Government and Politics: Market-Driven Viability vs. Neighborhood-Driven Viability.* We found that small, local business owners have strong local ties to the neighborhood where they build their businesses. However, market-driven viability creates barriers for local business owners to address community needs. Residents viewed the built environment as a reflection of public official priorities, pointing out differences between neighborhoods as an indication of which regions were prioritized with opportunities and which were not. Business owners shared

stories of how their neighborhoods have been neglected by the City (B3, B2). One business owner shared the stark differences between neighborhoods that were over-developed with national brands versus neighborhoods where businesses were primarily locally owned (B4).

Business owners' ties to the community were a crucial component in their story of opening a business, with 4 participants self-identified as lifelong residents of the City. Addressing neighborhood needs became a powerful motivator for business owners. B3 expressed her desire to open a food truck for her neighborhood, a food desert, due to under-resourcing and historic marginalization. However, B3 never started the business, as the building department *"wanted to know a location, a day, and a time"* of the business despite the desire to provide food to service workers, which requires flexibility in the business location.

This example of a food desert in under-resourced neighborhoods connects with how market-driven viability prevents meeting community needs. A community representative of an underserved neighborhood long requested a grocery store, which was rejected since *"it was not viable from a market perspective"* (CM2). A community advocate discussed market viability as a question of livability: *"When I think of viability, I think it depends on capacity. For example, where I live, the closest supermarket to us is at least two miles away [...] I have to get in my truck with my wife and drive up a hill"* (CM3). This presents two different definitions of viability - viability for the market, and viability for a resident to live in a place. We see that local business owners are affected by the barriers to market-driven viability.

## 6 DISCUSSION

The roll-out of OneStop reflects the managerial priorities of standardizing the process while unintentionally exacerbating existing inequities and affecting the work for small, local business owners, especially "first-time applicants". In this section, we discuss how our findings relate to the broader context of government services, the tensions between standardization, equity, and equality, as well as design implications that lead to equitable practices.

### 6.1 Standardization, Equality, and Equity in the Public Sector

Public service innovation projects often cite equity and fairness as important goals [5, 37, 76, 81]. Equity means that individuals receive the specific resources and opportunities they need to be successful and thrive in the community [36, 38, 56]. Equity differs from equality in which everyone receives the same resources and opportunities regardless of their needs. Within public services, inequity is often framed as uneven service delivery regarding how resources are allocated and who receives the services [17]. Scholars have addressed inequitable public service distribution in urban development [12, 32, 54, 64] due to public sector employees' discretionary power [52] and politics [31]. For example, frontline workers, also known as street-level bureaucrats, adhere to the law while making independent, discretionary interpretations of the policies [2, 3, 48]. The residents' participation in local politics contributes to how public services are prioritized, especially as historically underserved communities have been discriminated against urban government departments [10, 31].

In our work, the technology transition proceeded with a seemingly generic focus on operational efficiency and fairness that involves treating every applicant the same, even when their needs could be quite different. The design of OneStop lacked addressing the needs of "first-time applicants" who had strong connections to their local communities. OneStop facilitates a form of datafication that allows the city to determine what categories are important for the department. However, the digitization of the application process hid structural disadvantages for applicants who then needed to seek external support through personal connections with politicians.

OneStop created a standardized interface for all applicants to receive a uniform service. However, we found that OneStop created tension between standardization and equity. People arriving there did not come with the same knowledge, flexibility, and resources to absorb unexpected delays or changes in their business plans or relationships with experts. "Frequent flyers" who had prior relationships with the building department relied on those additional rounds of communication to bypass any possible frustrations in the permitting process. "First-time applicants" experienced challenges in gaining customization or face-to-face interactions needed to interpret unique aspects of the permitting process.

OneStop standardized the work process for employees, allowing supervisors to monitor their work productivity. Our findings emphasize management procedures in the public sector, where labor reflects efficiency [46, 55] and labor output represents the reduction of variability [65]. Such managerial visions embedded in technology align with prior literature investigating the relationship between politics and technologies [26, 47].

### 6.2 Design Implication: Creating the Infrastructure for Equity Practices for Constituents

Scholars investigated ways to surface uneven service delivery and tackle inequitable practices through efforts such as civic engagement [41] and the surfacing of impacted stakeholders' voices [11, 30, 80]. Constituents who are well-informed have the skills to politically influence their community. This practice becomes an equity issue because it is through engaged citizens that political pressure is realized [31]. Researchers investigated strategies and tools to help citizens participate in political decision-making [41]. Building on prior research, we suggest design implications of early engagement of stakeholders in building equitable public technologies, as well as collecting disaggregated data and creating auditing infrastructure.

*6.2.1 Early and Meaningful Engagement of Constituents in the Design Process of Building Equitable Public Technology.* Our findings revealed a contradiction between the goals of the technology deployment and the wants of local business owners. The building department saw the implementation of OneStop as a success due to the increase in the volume of applications and the efficiency of the work. However, local businesses desired to have dialogues about the public process and their needs, which was difficult when the interpersonal interactions became hidden behind a digital interface. They wanted the department to be involved in the success of their

businesses and have a positive impact on the neighborhood. We also identified a misalignment between the deployment of technology and the needs of local residents in the context of the building department. This misalignment has been found by prior scholars investigating the impact and perceptions of technologies used by the public sector.

Researchers discovered that the lack of engagement with stakeholders in the early stages impacted those whose needs and desires were not met [14, 44, 70, 72, 83]. It is after the technology has been developed that HCI is brought into the picture to investigate current challenges and the impact of public technologies [11, 42, 72, 80]. For example, Stapleton et al. investigated impacted stakeholders in the context of child-welfare advocates for no-tech or low-tech alternatives in high-stakes decision-making [80]. Robertson et al. found that the school matching algorithm intended to encode equity did not align with parents' real needs and desires [72].

Hence, we suggest that public sector agencies bring impacted constituents to the early stages of the design process of building or adopting new technologies. Prior work has investigated the need for involving stakeholders in the early stages of designing technologies [19, 20, 29, 78]. Expanding on past literature, we suggest that public sector agencies balance the misalignment of priorities between public sector agencies and communities before harm is inflicted upon the constituents. We identified a clear link between the neighborhood that the business resides in and the business owners' desires to improve the neighborhood, often addressing the gaps they saw in their community. The relationships residents have with their communities make them the experts in identifying collective needs and driving values. Hence, we advocate for researchers and practitioners in public technologies to focus on giving residents the power to address their collective needs (access, agency, resources) [16, 23]. This suggestion echoes calls by other HCI scholars to employ tactics to support community efforts [15, 82].

*6.2.2 Auditing Infrastructure and Disaggregated Data.* Auditing the performance of a public sector department is a way to assess equity and compare its service delivery across applicants. When we first started this study, we aimed to do an audit in addition to interviews with different stakeholders. However, this goal was not achievable since the publicly available dataset lacked essential variables and information needed to understand the equity component of the building department services, such as length of the application process, delays, and whether the business was minority-owned. This auditing attempt highlights a need for the public sector to develop data auditing infrastructure, including an evaluation of what data is needed for a thorough analysis of equitable public service delivery. The auditing infrastructure should not only be flexible enough to encapsulate an overview of the City function [30] but also address concerns regarding data privacy [79] and public sector's misuse of personal information [71].

As a means for auditing, we suggest the building department collect disaggregated data, data broken down into smaller subgroups such as race or gender [33]. Disaggregated data enables detailed analysis of discrimination and disparate impact based on different subgroups [49, 51, 64]. Hence, we suggest collecting data on local needs, reflecting the meaning of viability for residents.

## 7 CONCLUSION

In this paper, we interviewed public sector employees and constituents to understand how the City's building department transitioning to a digital service impacts the work of these groups. Surfacing the work practices impacted by technology led to findings of inequitable practices that were exacerbated by OneStop, which reduced the two-way communication between public sector employees and their constituents. Our findings highlight that managerial priorities towards standardization contributed to the integration of technologies in the building department, creating a misalignment between the public sector and its constituents. Based on these findings, future design implications to create infrastructure for equitable practices are necessary. With technology transition standardizing the work process for employees and impacting the work for constituents, future design could investigate early and meaningful engagement of constituents when designing public technologies and creating an auditing infrastructure to assess equity.

## ACKNOWLEDGMENTS

We thank all our participants for their time and input in this research. We appreciate the City's building department and the local community for sharing their experiences and knowledge. We thank our anonymous reviewers for their feedback, which helped improve this draft. This work has been partially supported by the National Science Foundation (NSF) under Award No. 1952085, 2129038, and the Carnegie Mellon University Block Center for Technology and Society Award No. 55303.1.5007719.